\documentclass{aastex}
\usepackage{emulateapj5}
\usepackage{onecolfloat5}

\input epsf
\def\plotone#1{\centering \leavevmode
\epsfxsize= 1.0\columnwidth \epsfbox{#1}}

\newcommand{\gsim}{\;\rlap{\lower 2.5pt
 \hbox{$\sim$}}\raise 1.5pt\hbox{$>$}\;}
\newcommand{\lsim}{\;\rlap{\lower 2.5pt
   \hbox{$\sim$}}\raise 1.5pt\hbox{$<$}\;}
\newcommand{\be}{\begin{equation}}
\newcommand{\beq}{\begin{equation}}
\newcommand{\ba}{\begin{eqnarray}}
\newcommand{\ee}{\end{equation}}
\newcommand{\eeq}{\end{equation}}
\newcommand{\ea}{\end{eqnarray}}

\newcommand{\bea}{\begin{eqnarray}}
\newcommand{\eea}{\end{eqnarray}}
\newcommand{\bean}{\begin{eqnarray*}}
\newcommand{\eean}{\end{eqnarray*}}

\begin{document}

\twocolumn[
\title{CMB Signatures of Extended Reionization}
\author{Lloyd Knox}
\affil{Department of Physics, University of California, Davis}

\begin{abstract}
I comment on the challenges and opportunities presented by
an extended period of reionization.  
\end{abstract}]

\section{Introduction}

The argument for an extended period of reionization is as follows.
The WMAP has detected the correlation between temperature and polarization
on large angular scales \citep{kogut03} that has an amplitude proportional
to the 
total optical depth of CMB photons to Thomson scattering, $\tau$
\citep{sunyaev80,
zaldarriaga97a,kaplinghat02}.  Modeling
reionization with a single sharp transition at $z_{ri}$, a
multi--parameter fit to the WMAP data gives $z_{ri} = 17 \pm 5$
\citep{spergel03}.
On the other hand, the evolution of quasar spectra from $z=6.3$ and
$z=6.4$
to $z = 6$ shows a rapid decrease in the amount of neutral Hydrogen,
indicating
the end of reionization \citep{fan03}.  A simple
interpretation to explain these two very different datasets is that
reionization started early, $z_{ri} \sim 20$, but did not conclude
until much later ($z \sim 6$). 

An extended period of reionization has its effect at low $\ell$ in
the polarization (as detected by WMAP), and also at higher $\ell$ in
temperature.  In Section II we disucss the low $\ell$ effect and
in Section III the high $\ell$ effect.  In section IV we discuss
them in combination.  All the work I present here is described in
more detail in \citet{haiman03} (reionization models), 
\citet{kaplinghat03a,holder03} (low $\ell$ signal) and \citet{santos03}
(high $\ell$).

\section{The Reionization Bumps}

\begin{figure}[htbp]
\begin{center}
\vspace{-0.2cm}
\plotone{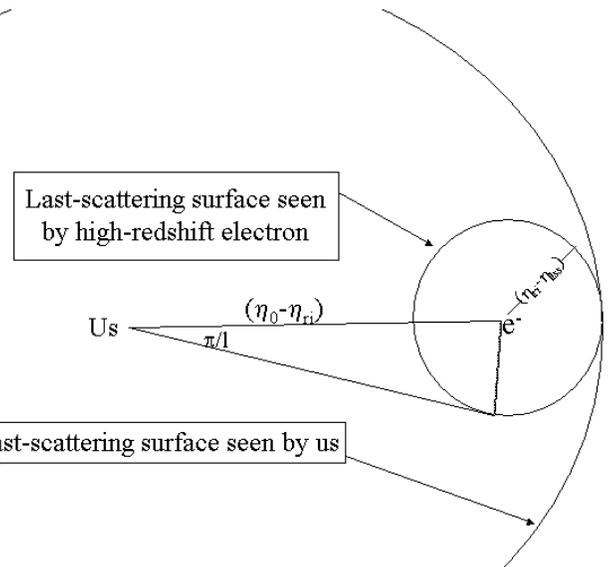}
\vspace{-0.6cm}
\caption{Temperature quadrupole generated by projection of monopole variation
on the electron's last--scattering surface generates linear polarization
on large angular scales today.
}
\label{fig:bump} 
\end{center}
\end{figure}

\begin{figure}[htbp]
\vspace{-0.2cm}
\plotone{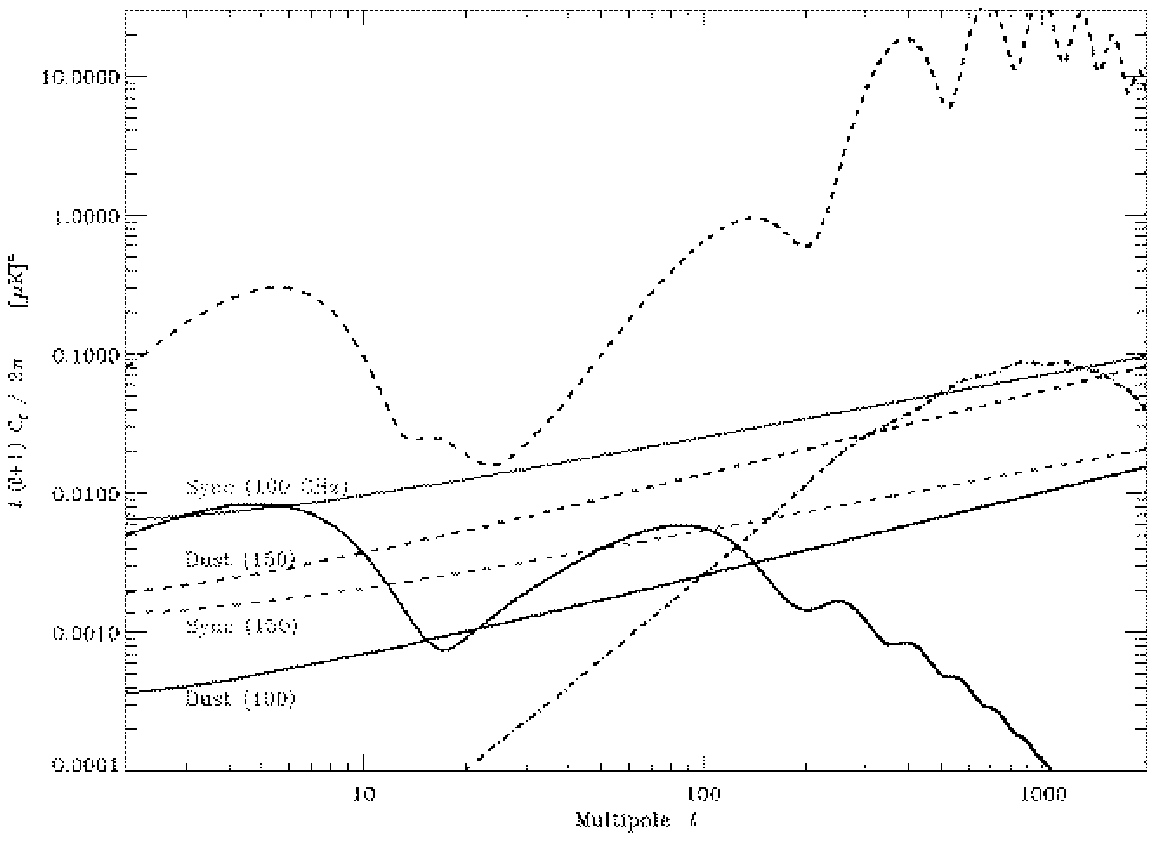}
\vspace{-0.6cm}
\caption{
CMB power spectra (unlabeled) and foreground spectra
(labeled).  The CMB power spectra are for $\tau = 0.17$ and $T/S =
0.28$.  E mode power spectrum (dashed), tensor CMB B mode power
spectrum (solid) and scalar B mode power spectrum (dot-dashed). Figure
courtesy Eric Hivon.  Foreground curves from \citet{prunet98} (dust)
and \citet{giardino02} (synchrotron).
}
\label{fig:foregrounds} 
\end{figure}

Quadrupole radiation incident upon an electron leads to linear
polarization.  For a review of CMB polarization, see \cite{hu97a}.
As drawn in Fig.~\ref{fig:bump} a free electron
at high redshift sees a quadrupole from free--streaming of the
monopole on its last--scattering surface.  A monopole
with wavenumber $k$ free streams primarily into $l=k(\eta_{ri} - \eta_{lss})$
so the quadrupole is given primarily by $k=2/(\eta_{ri} - \eta_{lss})$.
The linear polarization thus generated at $\eta_{ri}$ projects to 
$l = k(\eta_0-\eta_{ri}) = 2(\eta_0-\eta_{ri})/(\eta_{ri} - \eta_{lss})$ 
today.  Thus, the feature appears at low $\ell$.  The polarization
signature is proportional to the amount of 
scattering and hence the optical depth $\tau$.  Therefore
$C_l^{EE} \propto \tau^2$ and $C_l^{TE} \propto \tau$.  See
\citet{zaldarriaga97a} and \citet{kaplinghat03a}.  The same
considerations lead to a reionization bump in the tensor B mode with
$C_l^{BB} \propto \tau^2$ so higher $\tau$ can improve the detectability 
of gravitational waves as quantified in \citet{knox02,kaplinghat03b}.
The polarization angular power spectra are shown in Fig.~\ref{fig:foregrounds}.

The low l polarization has already been detected by WMAP through
the correlation of this effect with the temperature map \citep{kogut03}.  
From the WMAP measurements only one number can be inferred:  a joint fit of
the TT and TE power spectra to a six-parameter model results in
$\tau = 0.17 \pm 0.06$ \citep{spergel03}.  If  the EE power spectrum is measured with
cosmic variance precision, there are 5 uncorrelated numbers (the amplitudes of
5 eigenmodes of the ionization history), that can be measured with
signal-to-noise greater than 1 \citep{hu03}.  These 5 numbers will provide
strong constraints on models of the first generation of stars,
since these are presumably what cause the reionization of the
inter-galactic medium.

Models of foreground polarization indicate that this low l signal
can, in principle, be measured with near cosmic variance
precision, as seen in Fig.~\ref{fig:foregrounds}.  Although these
models are highly uncertain.  We will know more soon from further
releases of WMAP data.

To quantify how much improvement is possible beyond WMAP we show
forecasted constraints on $\tau$ and the initial reionization redshift
$z_e$ assuming statistical errors only for WMAP, Planck/LFI and
Planck/HFI.  Model B has sudden and complete reionization at $z_e=15$.
Model A has a two--stage reionization, first increasing $X_e$
suddently to 0.42 at $z_e=25$ and then suddenly completely ionizing at
$z=6.3$.  The ionization fraction, $X_e\equiv n_e/n_p$ cannot be greater 
than 1.08 if Helium does not doubly ionize.  Thus for a given $z_e$ there is a
minimum value of $\tau$ that gives rise to the straight lines in the
lower portions of the contours around model A.

\begin{figure}[htbp]
\vspace{-0.2cm}
\plotone{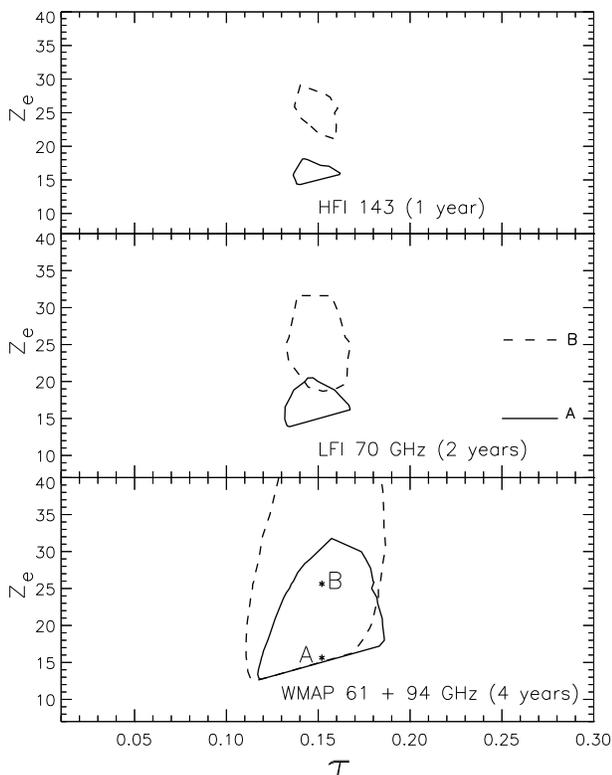}
\vspace{-0.6cm}
\caption{
Forecasted 68\% confidence levels for reionization parameters in a 
two-parameter model.  Although WMAP sensitivity is insufficient to distinguish
models A and B, Planck sensitivity is sufficient.
}
\label{fig:2param} 
\end{figure}

\section{The kSZ power spectrum}

An extended period of reionization (as suggested by combination of
WMAP data and high-redshift quasar spectra) is likely to be a
period with a highly inhomogeneous reionization fraction.  Prior to
percolation of the ionized regions, the
ionization fraction will be near unity in the vicinity of the
sources of the reioinizing radiation and zero far away from any
sources.  If the sources are in high-mass halos, as semi-analytic
models suggest, then the patches of high ionization fraction will
be highly correlated.  These correlated patches can lead to a
kinetic SZ power spectrum with an amplitude of about $10^{-12}$ at $l$
values of 1500 and higher.  At $l > 3000$ this kinetic SZ power
spectrum may be the dominant source of flucutation power on the
sky, for components with the same spectral signature as thermal
fluctuations about a 2.7 K black body.  Here we reproduce a figure
from \citet{santos03} showing a range of the kSZ power spectra that
come from the reionization models of \citet{haiman03}.

\begin{figure}[!bt]
\vspace{-0.2cm}
\plotone{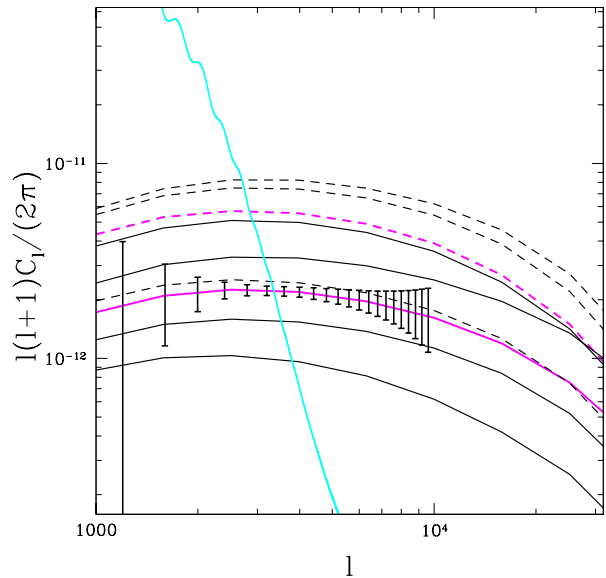}
\vspace{-0.6cm}
\caption{Patchy power spectrum for a variety of reionization models.
Solid lines correspond to the following values of $\tau$ (from
bottom to top): 0.05, 0.11, 0.17, 0.24, 0.31. Dashed lines correspond
to $\tau=$0.11, 0.17, 0.22, 0.28. Also shown are the expected
measurement errors and primary lensed CMB power spectrum.}
\label{fig:models} 
\end{figure}

The error bars on one of the curves are for an observation of
1\% of the sky with 0.9' resolution.  They are due to the
``noise'' from primary + lensing CMB and residual point sources.  
Instrument noise is assumed to be subdominant.  We assume
thermal SZ has been removed by taking advantage of its unique
spectral dependence.

We see that the models do not show much variation in shape, but
only in amplitude.  There is a degeneracy between the clustering bias of
the dark matter halos hosting the reionizing sources and the 
optical depth.  The low $\ell$ measurements can break this degeneracy,
allowing us to place some constraint on the clustering bias, and
therefore the halo masses.

Although predicting the amplitude of the signal is difficult, the
shape appears to be robust, particularly at $l < 3000$ where the
kSZ power spectrum might contaminate attempts to measure cosmological
parameters and reconstruct the gravitational lensing potential.  We
can therefore model this contaminant with one free parameter:  an
amplitude.  Doing parameter estimation without such modeling (i.e.,
ignoring the kSZ power spectrum) can lead to significant biases
for parameter estimation from Planck and higher--resolution observations
\citep{knox98,santos03}.
But \citet{santos03} find that modeling the kSZ power spectrum as
this robust shape times a floating amplitude removes all significant
parameter estimation biases even for an experiment that is cosmic--variance
limited out to $l=3,000$.

The simple prescription for removing the bias works because there is
so little variation in the shape of the kSZ power spectrum in our 
reionization models.  This shape is merely a projection of the matter power
spectrum from high redshift.  The small dependence on shape that there
is comes from the different mean angular diameter distances in the different
models.  If necessary, the modeling could be extended to include this
as a free parameter also.

If the real kSZ power spectrum shape (at $l < 3,000$) were  
significantly different from the shapes we calculate then
our simple modeling (with one or maybe two parameters) may not
be sufficient.  However, this could only be the case if the 
density of free electrons does not trace the density of matter
on scales larger than about 3 Mpc.  Note that even if $X_e$ were
completely homogeneous, the density of free electrons traces
the density of matter.  Even in this homogeneous $X_e$ case the
shape would still be a projection of the matter power spectrum,
although this time from a smaller mean angular-diameter distance.

Our calculations have relied on semi--analytic models of reionization.
For a more numerical approach, see the contribution from 
Naoshi Sugiyama.  For more on patchy reionization, see 
\citet{aghanim96,gruzinov98,knox98,valageas01} and other papers already
mentioned above.

\acknowledgments{I am deeply indebted to my collaborators Asantha
Cooray, Zolt\'an Haiman, Gil Holder, Manoj Kaplinghat, Chung-Pei Ma, 
Mario Santos and Yong-Seon Song
and to NASA who have supported this work through grant NAG5-11098.  I
also thank Andrew Lange, Charles Lawrence and 
Martin Rees for useful conversations.  }

\bibliography{reion}

\end{document}